# Non-Enzymatic Glucose sensing properties of NiO nanostructured flower decorated Exfoliated Graphite Electrodes

Piyush Choudhary[1], Chhavi Chetiwal[1], Chandra Prakash[1], Vijay K. Singh[2], Ambesh Dixit[1,3,*]

[1]*Advanced-Mateirals and Device (A-MAD) Laboratory, Department of Physics, Indian Institute of Technology Jodhpur, Jodhpur, Rajasthan, 342030, India.*
[2]*Department of Physics, Indian Institute of Technology, Ropar, Punjab, India.*
[3]*Rishabh Centre for Research and Innovation in Clean Energy (RCRICE), Indian Institute of Technology Jodhpur, 342030, India*
*Corresponding Author Email: \*ambesh@iitj.ac.in*

**Abstract:**

Nanostructured transition metal oxides (TMO) are extensively explored materials for non-enzymatic glucose sensors. TMOs such as Iron oxides( $\alpha$-$Fe_2O_3$, $\gamma$-$Fe_2O_3$, $Fe_3O_4$, etc.), NiO, CuO, $Cr_2O_3$, etc. have been utilized as electrocatalysts for glucose determination. Tremendous efforts have been put into identifying the impact of different morphologies of these materials on the glucose-sensing performance. The larger surface area of the flower and wire-shaped catalysts make them better performing amongst other morphologies. Interestingly, it is important to note that most of such studies are on standard Glassy Carbon electrodes. Further to enhance the Electrochemically active surface area (ECSA) of the electrode, Carbon nanomaterials such as reduced Graphene Oxide (r-GO) and Carbon Nanotubes (CNTs) are used as additives. Exfoliated Graphite paper electrodes offer better electrochemical characteristics than GCE electrodes due to their much larger ECSA. This study presents the non-enzymatic glucose sensing properties of NiO nanoflower-decorated Exfoliated Graphite electrodes. The amperometric detection of glucose shows a linear increase in current over a physiologically relevant wide range of 0-10 mM. The electrodes offer a better sensitivity of 304.12 $\mu A\ mM^{-1}\ cm^{-2}$ and a Limit Of Detection (LOD) of 100 $\mu M$. In addition, the electrodes showed high selectivity towards glucose in the presence of other interfering species such as Ascorbic acid, Fructose, Sucrose, and NaCl.

**Introduction:**

Diabetes mellitus is one of the most critical global health challenges. This modern-day silent health catastrophe is a chronic metabolic disorder caused by an elevation in levels of blood glucose. According to recent estimates by the International Diabetes Federation (IDF), approximately 783 million people are predicted to suffer from diabetes globally by 2045, as predicted in the 10$^{th}$ edition of the IDF Diabetes Atlas[1]. This enormous figure is expected to increase drastically to 853 million by 2050, according to the 11$^{th}$ edition of the IDF Diabetes Atlas[2]. Astoundingly, over 4 million diabetes-related deaths are estimated annually. It is a cause of alarm that nearly half of all adults with diabetes remain undiagnosed[1], [2]. This forecast presents a massive market for glucose sensors. The global demand persists in the near future as no permanent cure for the disease exists. The glucose sensor is a billion-dollar market product dominating the field of commercial biosensors. At present, commercial glucometer brands such as Abbott[3], LifeScan[4], Bayer[5], and Dexcom[6], while reliable, typically depend on enzyme-based systems and come with high recurring costs. A comparative study of different glucometers to test performance and cost parameters was done by S. Jeyaram[7]. According to the study, a single test strip costs around ₹30–₹60, and glucometers typically cost around ₹1500. Noticeably, advanced continuous glucose monitoring (CGM) devices such as the Dexcom G6 range from ₹15,000 to ₹20,000[6]. The study presents that all the commercial glucometers used are dependent upon enzymes. The enzymatic glucometers suffer from inherent drawbacks, as can be easily interpreted from the literature. As a result, non-enzymatic glucose sensing is a domain of peaked interest in the biosensing research community. The non enzymatic glucose sensing systems avoid biological enzymes to perform direct electrochemical glucose oxidation on their conductive surfaces. Therefore, the mechanism presents two important requirements for development of non-enzymatic glucose sensing systems. First, excellent electrocatalytic activity that can be used for stable and efficient bioreceptor (oxidizing agent) for glucose. Second, superior electrical conductivity that can help in better transfer or transduction of charges (signal) due to the electrocatalyst. Apart from the two requisites, the important matter of weight is the minimized economic burden of the testing device. The affordability and desired utilization of glucometers is restricted by high cost of the non-reusable units (test strips). With the increasing population in India, the challenge is to provide Point of Contact (PoCt) detection devices to its low- and middle-proceeds class at almost trivial cost. This brings forth the significantly substantial problem, fabricating a state-of-the-art glucose sensor with high cost-efficiency without compromising the sensitivity or even surpassing the performance of present glucose sensors.

The fulfillment of first requirement of the problem is taken care of using several materials such as Metal Nanoparticles (MNPs)[8], Transition metal oxides (TMOs) [9], M-Xenes [10], Metal Sulphides, and most recently Metal organic Frameworks (MOFs)[11]. TMOs are the most preferred choice for glucose oxidation due to their excellent electrocatalytic activity, easy synthesis, and far lower economic input. The availability of multiple stable oxidation states allows TMOs to be suitable for glucose oxidation. Several materials, such as Iron oxide[12], [13], Cobalt oxide[14], Copper oxide[15], and others have been tested rigorously over the past six decades for this purpose. However, nickel and its oxides have been the pioneers for non-enzymatic glucose oxidation. In fact, the first non-enzymatic glucose oxidation properties were put forth by Fleishmann et. al. using Nickel electrodes in 1971[16]. Nickel oxide (NiO) is one of the superior materials for glucose sensors because it demonstrates exceptional electrical conductivity alongside chemical stability in basic medium and high catalytic abilities for glucose oxidation[17]. Its abundance on earth makes it a much more cost-effective option. It is often referred to as the destitute sibling, belonging to the same group as Pt and Pd in the periodic table. Non-enzymatic glucose sensors utilize different nanostructures, including hierarchical flake-flower[18], [19], nanosheet[20], [21], and nanowires [22], [23]. Amongst all the studied morphologies, flower-like nanostructures possess extensive surface space that increases active interaction sites for glucose molecules, making them more sensitive and effective than others. The NiO nanoflowers are therefore one of the most obvious choices as they tick all the boxes.

Most electroactive materials typically lack electrical transduction properties. Thus, assistance in the conduction of charges generated during glucose oxidation is the other important requirement for better performance. The maximization of electrochemical performance of NiO nanoflowers requires use of a substrate that can complement the electrocatalytic properties of NiO with superior charge transfer. The electrochemical sensing traditionally relies heavily on the Glassy Carbon Electrode (GCE) as a working electrode. The GCE demonstrates excellent electrical conductivity, minimum background noise, and good chemical inertness[24], [25]. That's why literature is full of articles related to non-enzymatic sensing that use GCE as their base electrode. Despite the advantages, GCE faces practical limitations such as high cost, considerably large size, and non-disposability. The other common substitutes designed for overcoming these limitations are screen printed electrodes (SPE) and ITO/FTO-coated glass. These substitutes superiorly overcome the shortcomings of the GCE as a PoCt device. However, to enhance the performance, all the mentioned substrates require modification using nanomaterials such as r-GO, CNT, and other carbon nanomaterials (CNMs)[26]. A better substitute, in fact, can be the use of these nanomaterials directly as substrates rather than modifiers. Exfoliated graphite (Ex-G) is a carbon material having a highly porous structure and very high surface area. It can be easily converted into a desired shape by compression under mechanical pressure. This property helps fabricate flexible sheets that can be cut into a desired shape and size to act as a substrate. It is thus one of the most viable candidates amongst the group of CNMs. Ex-G has recently been established as a potential substrate due to its larger electrochemical area than GCE and SPCE. These electrodes offer better sensitivity with less modification required for better transduction. Moreover, it fits exceptionally well with the economic aspect as the material is highly cost-efficient, and the materials and methods needed for its electrode fabrication are also simple. The Ex-G electrode can therefore be tested for improved performance against the standard GCE as the working electrode substrate. Despite being used as a substrate in some studies in literature[27], the electrode fabrication of Ex-G has also been a little challenging. Often, research groups try to emulate the GCE for its traditional design. We have chosen a simple and more robust design in terms of fabrication. The study aims to design these Ex-G electrodes and their apt modification using the NiO nanoflowers as an efficient yet cost-effective glucose sensing device.

**Experimental:**

**Materials:**

Natural graphite flakes, $H_2SO_4$, $HNO_3$, $KMnO_4$, $NiCl_2 \cdot 2H_2O$, Polyvinyl pyrrolidinone (PVP), Urea, Ethanol 99.9%, Whatman Filter Paper, Copper wire, Silver paste, Kapton Insulating tapes, Aeraldite Epoxy Glue, Glass slides, De-ionized Water (DI Water), NaOH. The electrochemical experiments were performed using Palmsens potentiostat, and a standard three-electrode process was utilized for the study. The reference electrode was Ag/AgCl (3M KCl), and a Platinum wire electrode being used as the counter-electrode, was purchased through Palmsens.

**Methods:**

**Synthesis of Exfoliated Graphite (Ex-G):**

A modified method for easy exfoliation of natural graphite flakes (NGF) was adopted, which has been previously reported by our group[28]. Herein, $KMnO_4$ was used to enhance the rate of intercalation. In a traditional method 3:1:1 ratio of $H_2SO_4$ : $HNO_3$ : Graphite is taken. Our adopted method adds 1:0.44 ratio of Graphite : $KMnO_4$ to the process. So, a 1 g sample of NGF was dipped in a concentrated acid solution of 28 mL of $H_2SO_4$ and 9 mL of $HNO_3$. This mixture is stirred for an hour and kept in an ice bath to prevent oxidation of graphite. Under continuous stirring in an ice bath, 0.44 g of $KMnO_4$ is slowly added to this mixture to avoid an exothermic reaction. Adding a small quantity of $KMnO_4$ in the process increases the intercalation rate due to the cleavage of the C-C bonds, cutting down particle size. The mixture is then left under continuous and slow stirring for 3 hrs. This particular step is responsible for the intercalation of $SO_4^{2-}$ and $NO_3^-$ ion species into the interplanar spacing of the graphitic sheets present in the flakes. The reaction is then quenched by the addition of excess DI water. The product is then washed multiple times with DI water to remove all the unreacted $KMnO_4$ and acids present completely. The acid-treated graphite flakes are termed Intercalated Graphite Flakes (IGF). The IGF is then subjected to microwave irradiation for expansion at a power of 800 W for 2 minutes to achieve exfoliated graphite (Ex-G). The worm-like, fluffy exfoliated graphite is then compressed under 20 tons of pressure for six hours to form a sheet, which is used further in experiments.

**Synthesis of NiO Nanostructured Flowers:**

Hydrothermal method is one of the most suitable synthesis methods that provides control over morphology of the nanomaterial. A method from literature[29] was reoptimized to synthesize NiO nanostructured flowers. A detailed schematic of the process is shown in Figure 1. Herein, 0.95 g of $NiCl_2 \cdot 2H_2O$ was dissolved in 20 mL of absolute ethanol. A ~3% PVP solution was prepared by mixing 0.6 g of PVP into 20 mL of DI water. A combination of ultrasonication and mechanical stirring was used to dissolve the PVP. Once completely dissolved, it was added dropwise to the Ni precursor under vigorous stirring. The rate of addition of PVP sometimes influences the morphology of the flower product. It is thus important to add it slowly. The solution is stirred for 30 minutes. Finally, a 20 mL 0.9 M urea solution is prepared in equal proportions of ethanol and DI water. This solution is added to the Ni precursor mixture and stirred for 10 minutes. The contents are then transferred to a Teflon-lined hydrothermal vessel and sealed in a stainless-steel autoclave. The autoclave is kept in an electric oven at 160 ºC for 10 hrs to obtain $Ni(OH)_2$ precipitate. The precipitate is washed with DI water and ethanol several times to remove excess byproducts. The precipitate is collected by centrifugation at 4000 rpm for 5 minutes. This precipitate is dried in the oven at 60 ºC. To convert the hydroxide precipitate powder into oxide, it is annealed at 450 ºC for 4 hrs. Thus, the final NiO nanostructured flower (NiO NF) is collected after annealing in a box furnace.

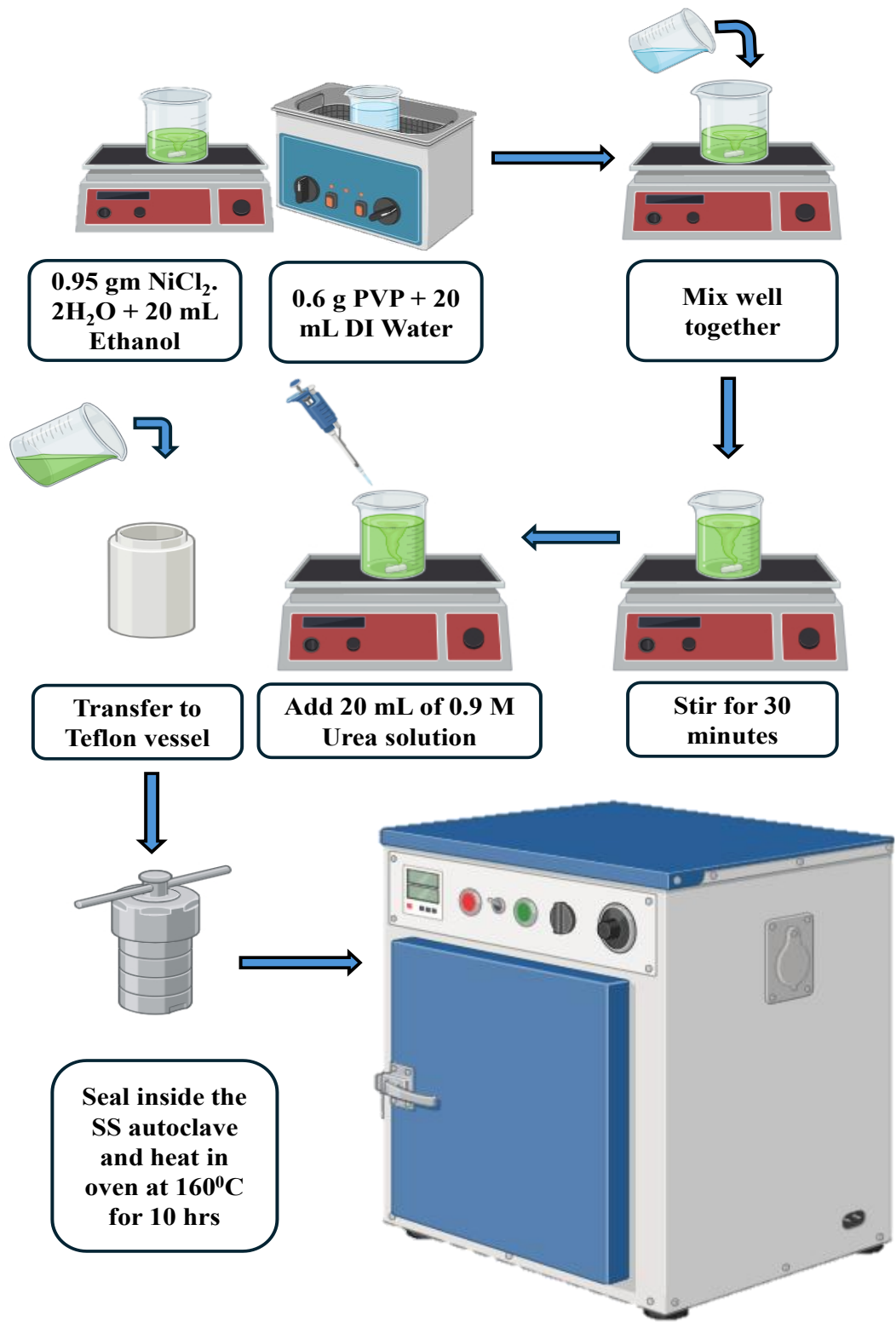

**Figure 1:** A schematic diagram for the workflow involved in the hydrothermal synthesis of NiO nanostructured flowers

**Fabrication of NiO NF@Ex-G electrode:**

The Ex-G sheets of thickness 0.5 mm are cut into 5 mm circular pieces using a punch. A copper wire is then attached to one of its flat surfaces using silver paste. This wire-attached Ex-G disk is then fixed on the surface of a cleaned glass slide using Kapton tape. The other orifices of the Ex-G electrode are then sealed with aeraldite epoxy glue. This is done to prevent copper wire or other conducting surfaces except for a single basal plane, from being in contact with an electrochemical solution. This assembly is termed a bare Ex-G electrode. These electrodes can be modified similar to Glassy carbon electrodes for glucose sensing. Figure 2 shows the schematic of this modification. Firstly, a dispersion of NiO NF is prepared by mixing a known quantity of it in DI water such that the concentrations range from 0.5 - 3 mg/mL. 20 µL of this dispersion is drop-cast on the surface of cleaned Ex-G electrodes. The electrodes are named NiO NF 0.5, NiO NF 1 , NiO NF 2 and NiO NF 3, respectively, depending upon the concentration of dispersion used. It is then left to dry overnight in ambient conditions. 10 µL of 0.1 wt% Nafion solution is then coated over the electrode as binder and dried for 30 minutes in ambient conditions. These electrodes are stored in a dessicator before final use.

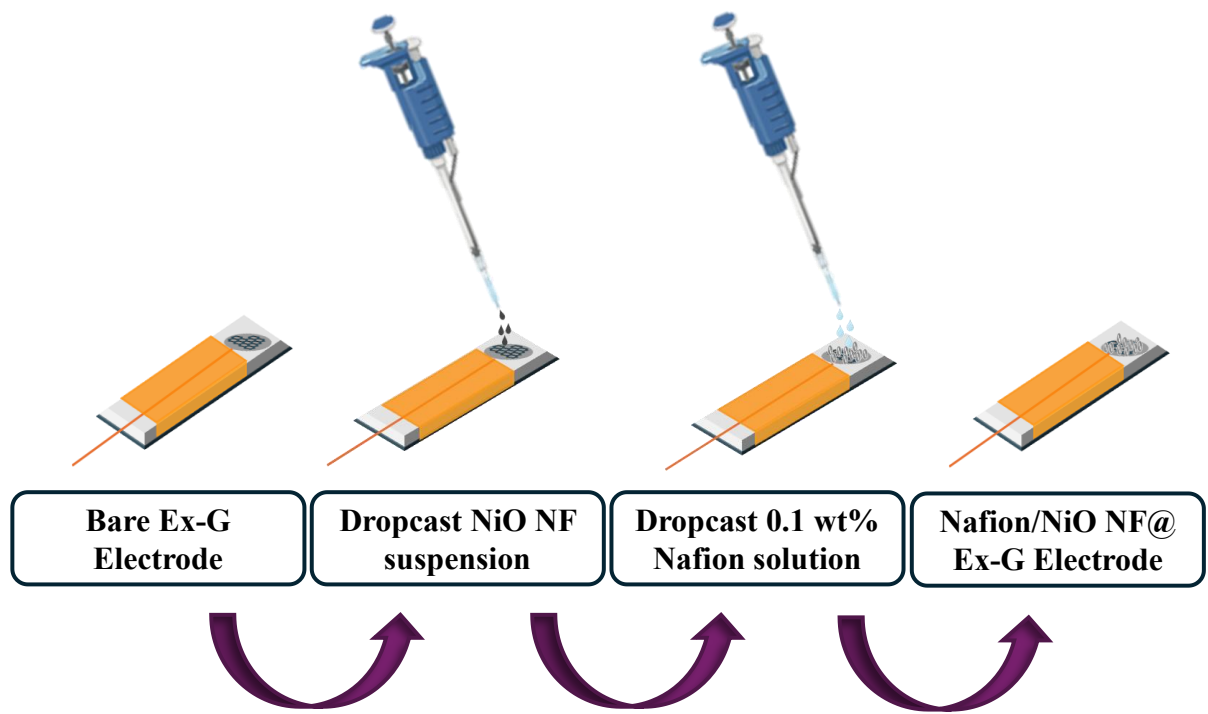

**Figure 2:** A schematic representing the steps involved in the fabrication of NiO NF@ExG electrodes.

**Electrochemical Detection Process:**

The electrochemical detection utilized 20 mL of 0.1 M NaOH as the analytical solution. All the processes were carried out using this solution only. A three-electrode setup was used for the detection process with Ag/AgCl (3M KCl) as the reference electrode and Pt wire as the auxiliary electrode. The working electrodes were fabricated as explained above. The cyclic voltammetric measurements were recorded from 0 to 0.8V at a 200 mV/s scan rate. Initially, 25 cycles of cyclic voltammetry were run for stability to achieve a stable baseline response. For glucose sensing experiments, chronoamperometry was run at 0.6 V with a time step of 0.01 seconds. To obtain a stable control response, the experiment was run for 100 seconds, then paused, and rerun after the addition of glucose stock solution. A 30-second analysis window was considered for analysis as most stable responses were achieved with this window.

**Results and Discussions:**

**Material Characterization:**

The detailed characterizations of the NiO nanostructured flowers and the NiO NF decorated Ex-G electrodes were carried out to analyze their structural, morphological, and functional characteristics. For the structural properties, X-Ray Diffraction and Raman Spectroscopy measurements were used. The functional groups present in the NiO NF were studied using FTIR spectroscopy. The morphological studies of the NiO NF and their decorated ExG electrodes were carried out using Field Emission Scanning Electron Microscopy. The X-Ray Diffraction is carried out using a Malvin Panalytical Powder X-Ray Diffractometer. The XRD data was collected within 15° - 80° 2θ range at $0.01^0$ step size and 12°/min scan rate. The detailed XRD pattern analysis of the NiO NF powder shows characteristic diffraction peaks, for example (111) at 37.20°, (200) at 43.24°, (220) at 62.33°, (311)  at 75.20°, and (222) at 79.83° 2θ, as shown in Figure 3(a) with green color diffractogram The peaks present in this XRD spectrum support the pure phase NiO formation without the presence of other oxide and hydroxide phases. The clear peaks (h, k, and l, all odd or even) confirm the face-centered cubic arrangement with the halite or rock salt NiO structure. The interplanar spacing for the most intense peak corresponding to (200) is calculated to be 0.21 nm, which matches closely with studies in literature [30]. The crystallite size determination was done using the Scherrer method. The crystallite or the grain size was ~16.34 nm. The nanosized domains present within the micron-sized flowers is the reason for calling them nanostructured flowers. A much clearer picture of this aspect is obtained with the morphological analysis discussed later in the section. The XRD spectrum of NiO NF decorated ExG was also recorded and is shown in Figure 3 a) (blue).

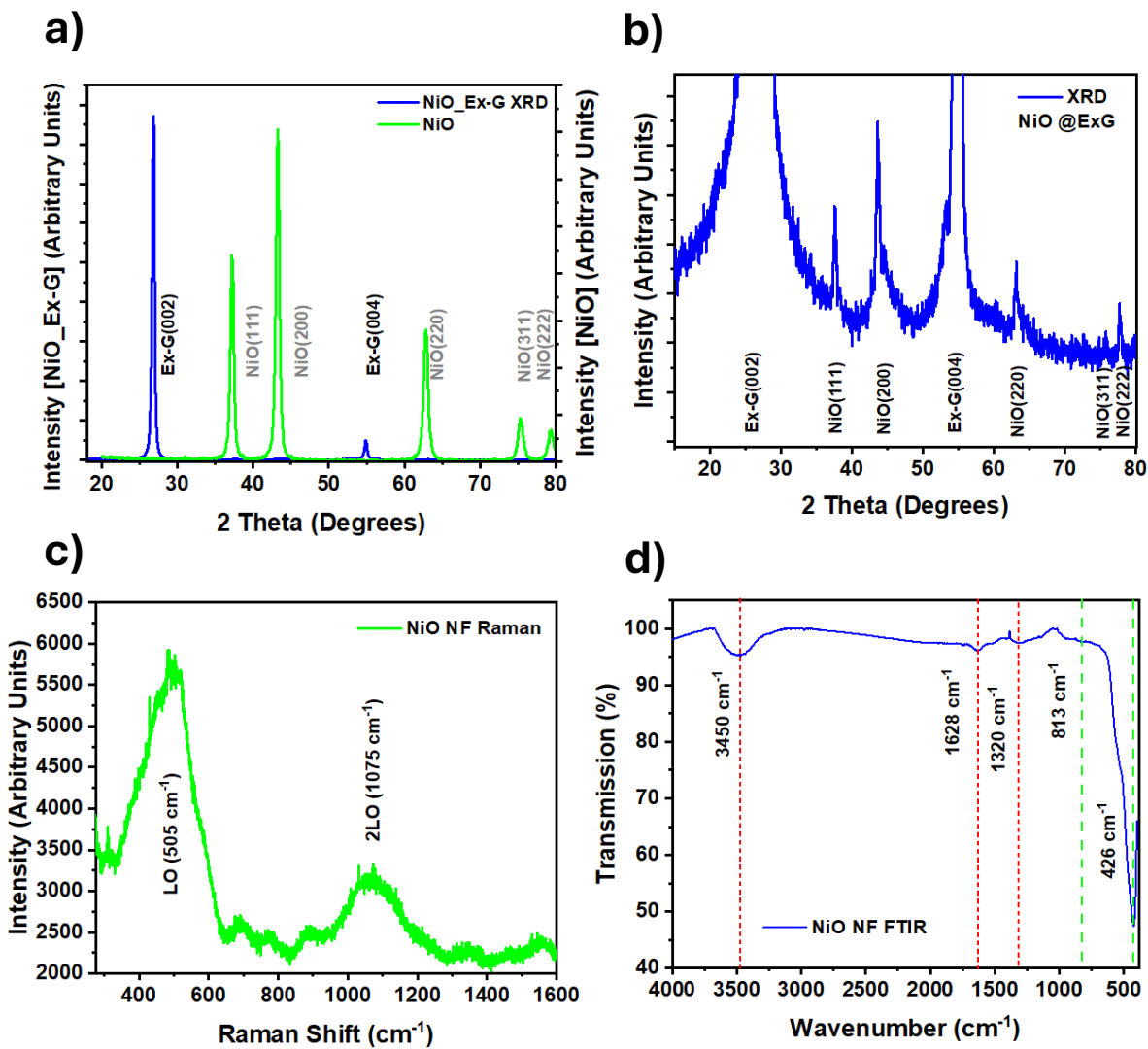

**Figure 3:** a) XRD spectra of NiO NF and NiO NF decorated GCE electrodes b) Enlarged XRD spectra of NiO NF decorated Ex-G electrodes (In Figure 3 a) showing presence of NiO. c) Raman Spectra of NiO NF decorated Ex-G electrodes. d) FTIR spectra of NiO NF powder.

The highly intense characteristic graphite peak at 26.5° corresponds to (002) plane, and its second-order diffraction is observed at 54.90°. The high intensity of these peaks caused the NiO peaks to be finitely submerged in the background. However, a closer look at the spectra confirmed the clear presence of NiO NF particles on its surface. An enlarged spectrum of the NiO NF decorated Ex-

G electrode is shown in Figure 3 b). It shows the presence of all five characteristic NiO peaks along with those of Ex-G.

Raman Spectroscopy was used to complement the structural analysis of XRD. The Raman spectroscopy of the NiO NF powder was carried out using a Horiba Labram Odyssy Raman Spectrometer at 1 mW fixed input power with 10 seconds of acquisition time. The spectrum collected using these parameters with a 20X scope is shown in Figure 3 c). The two $LO_1$ and $LO_2$ modes are distinctively present at 505 cm$^{-1}$ and 1075 cm$^{-1}$, respectively. These two peaks match well with literature confirming the formation of phase-pure NiO[31]. However, the TO mode usually present in the Raman signature of NiO is not clearly visible. This might be quite possibly due to the high background signals of the NiO NF powder. The nanostructured flower-shaped powder has high porosity, increasing the unwanted background signals.

With the structural confirmation using RAMAN and XRD spectroscopies, the functional group analysis was done using FTIR spectroscopy. A tabletop FTIR spectrometer was used in transmission mode for this purpose. The transparent pellets were prepared by adding small amounts of NiO NF powder to KBr powder, and data was collected from 400 to 4000 cm$^{-1}$ wavenumber range. The resultant FTIR spectrum of NiO NF powder is shown in Figure 3 d). The graph shows peaks at 426 cm$^{-1}$ for the fundamental NiO stretching, 813 cm$^{-1}$ for the Ni-O bending mode. These sync well with reports in the literature [32]. These peaks confirm the successful synthesis of phase-pure NiO. The peaks at 1320 cm$^{-1}$ for C=O stretching or $CO_3^{2-}$ are possibly present due to the oxidation byproducts of PVP (used as the capping and growth directing agent) at the surface of NiO NF. Finally, the peaks at 1628 cm$^{-1}$ for the bending mode of H-O-H and 3450 cm$^{-1}$ for O-H stretching are present due to NiO's surface adsorbed moisture and hydroxyl groups. Therefore, it is confirmed that no external impurities or functional groups are present in the sample. Apart from a few adsorbed carbon residues and moisture, the syntheized NiO NF material is phase-pure.

The most important characterization was the confirmation of morphology of NiO particles, which was done using FESEM. A low magnification FESEM image of NiO NF synthesized is shown in Figure 4 a). This shows uniform flowers closely resembling the marigold shape formed throughout the sample. For closer inspection, a high resolution FESEM image was collected, shown in Figure 4 b). The distinct, small, patterned petals of the flowers can be clearly seen in this image. Each petal is composed of small rice-grain structures of the order of a few nanometers. These have small pores between them, increasing the surface area of the petals and the overall structure. Thus, the image complements the XRD analysis of the grain or domain size of ~20 nm. These nanostructures present within the micron sized flowers of the NiO are the reason for calling them nanostructured flowers. The surface of the Ex-G electrodes was also inspected using FESEM. Figure 4 c) shows a low-resolution image of a polished and cleaned Ex-G electrode surface at lower resolution. The image shows wrinkled Graphene sheets compressed under pressure to form a flat surface. An image at a higher resolution is shown in Figure 4 d), which shows a polished, smooth surface without any surface indentations. The clean planar surface of the Ex-G before functionalization can be seen in this figure. A 3 mg/mL dispersion of NiO NF is drop cast on the surface, and nafion is finally coated for binding these particles onto the electrode surface. Finally, the NiO NF/Nafion-coated ExG surface at low and high magnification is shown in 4 e) and f), respectively. For better comparison, the magnification of the ExG and NiO NF coated Ex-G electrode is kept the same in 4 c) and e) as well as 4 d) and f). The last two FESEM images show the high surface coverage of Ex-G electrode by the NiO NF particles. Thus, by techniques of XRD, RAMAN, FESEM, and FTIR it is established that the NiO NFs are phase-pure and are uniformly coated onto the surface of the Ex-G.

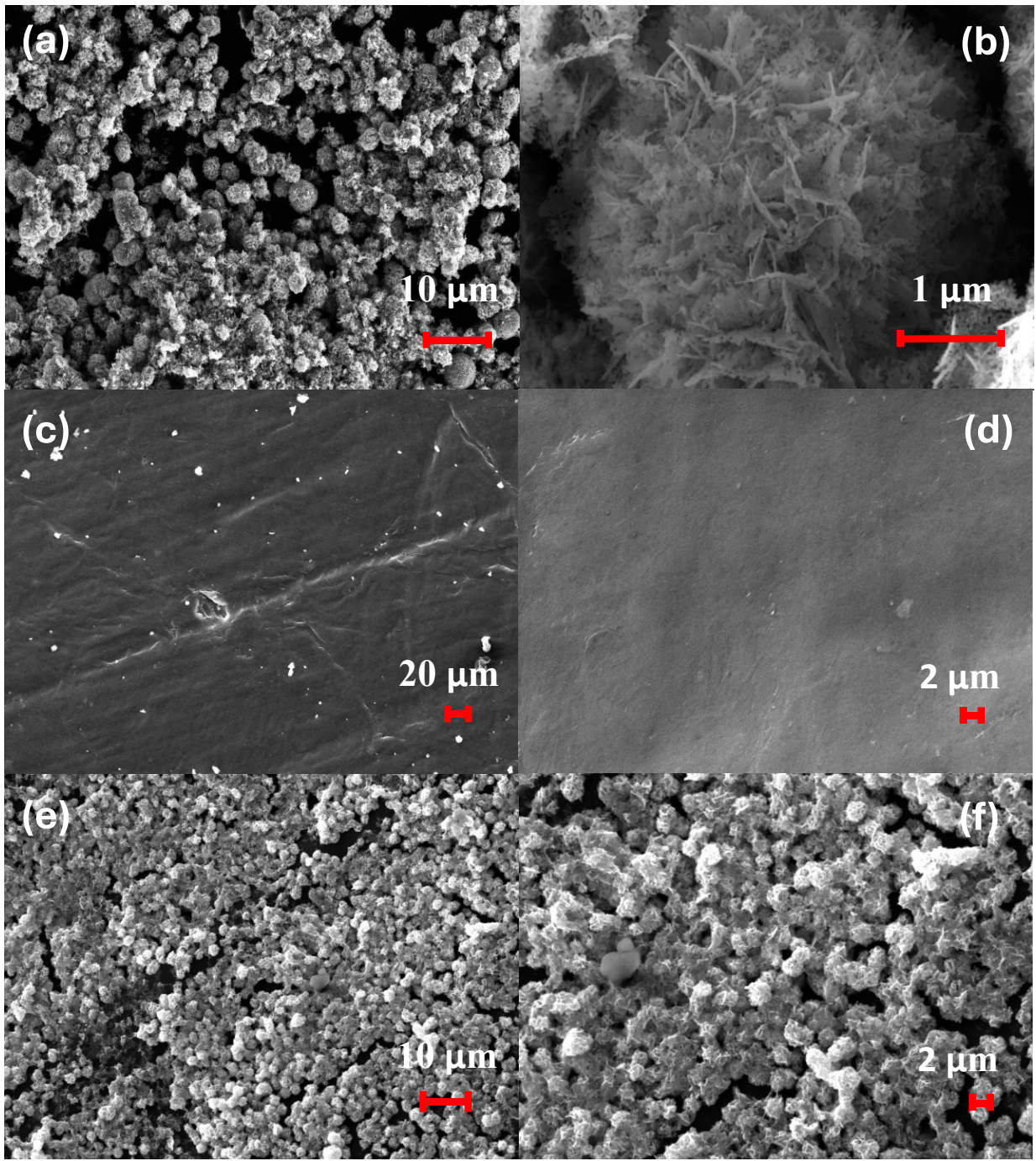

**Figure 4:** FESEM images of a) NiO NF at low resolution b) NiO NF at high resolution c) Ex-G electrode surface at low resolution d) Ex-G electrode surface at high resolution e) Nafion coated NiO NF 3 @Ex-G electrode at low resolution f) Nafion coated NiO NF 3 @Ex-G electrode at high resolution.

**Glucose sensing device characteristics and performance:**

The NiO NF@ Ex-G sensors fabricated as described in the previous section are characterized using electrochemical techniques for glucose sensing properties. This is initiated using Cyclic voltammetry in the presence and absence of glucose. Figure 5 a) shows the cyclic voltammograms of pristine Ex-G, NiO decorated Ex-G with and without glucose presence. The pristine Ex-G electrodes do not show any redox peaks within the potential window of 0 - 0.8 V. The NiO decorated Ex-G electrodes show peaks at 0.4 V and 0.55 V for oxidation and reduction, respectively. The $Ni^{+2}$ to $Ni^{+3}$ transition and vice versa in the basic medium have excellent reversibility. This transition mediates glucose oxidation. If the device is sensitive to glucose, then there is a considerable difference between the peak currents in the presence and absence of glucose. This difference is evident in the graph [Figure 5 a)], making it a good candidate for glucose sensing applications. In literature, the usual pathway for glucose oxidation with NiO in a basic medium is defined by the set of chemical equations given by Fleishmann[16]:

$$NiO + H_2O + OH^- \rightarrow Ni(OH)_2 + OH^-$$

$$Ni(OH)_2 + OH^- \rightarrow NiO(OH) + H_2O + e^-$$

$$NiO(OH) + Glucose \rightarrow Ni(OH)_2 + Gluconolactone$$

The basic medium turns the NiO into its hydroxide. This hydroxide compound upon electroanalytical cycling, is converted to a metastable state NiO(OH) which acts as an oxidizing agent for the glucose molecules. This is the most accepted mechanism for non-enzymatic glucose oxidation using NiO.

A scan-rate-dependent cyclic voltammetric experiment was conducted to gain further insight into the glucose oxidation process. This was done in the presence of 1 mM glucose in 20 mL of the electroanalytical solution. The graph of scan rate dependence of glucose oxidation process is shown in Figure 5 b). The results show an increase in peak currents and a shift in both peak potentials with an increasing scan rate. The increase in peak separation at a higher scan rates occurs due to lower reversibility of the redox reaction. It is important to point out that for reversibility of the electrochemical reaction, the peak separation should be smaller than 69 mV as predicted by the Nernst equation. In this case evidently, the reaction is completely reversible only at lower scan rates. When the peak currents (anodic and cathodic) are plotted against the square root of the scan rate, then they follow a linear trend. This confirms that Randles-Sevcik's equation is followed and that the process is diffusion-controlled. The plot of cathodic and anodic peak current vs the square root of scan rate is shown in Figure 5 c) and d) respectively. This plot shows that despite the linear nature, the fitting is not completely perfect as the $R^2 \sim 0.98$. This attribute may be because of adsorptive component in the mechanism.

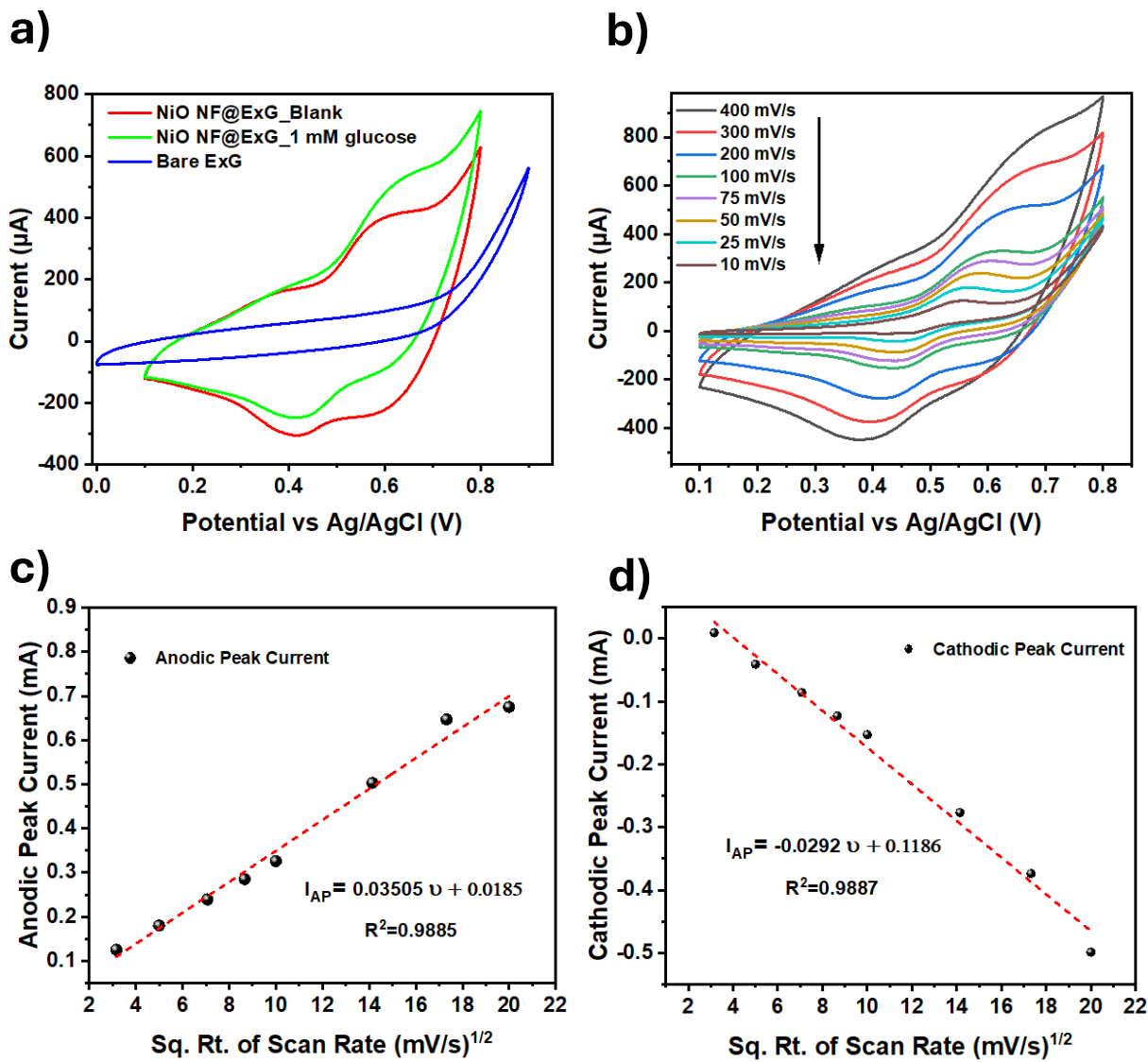

**Figure 5:** a) The cyclic voltammograms of Ex-G, and NiO NF@Ex-G in absence and presence of 1 mM glucose in 0.1 M NaOH. b) The scan rate dependent cyclic voltammograms of NiO NF@Ex-G electrodes in 1mM glucose + 0.1 M NaOH. c) Plot of square root of scan rate vs Anodic Peak current. d) Plot of square root of scan rate vs cathodic peak current.

Due to the highly porous nature of both the Ex-G and NiO NF, it is possible that the porosity is one of the factors contributing to the deviation from a linear nature. Thus, this deviation points out the adsorption of glucose onto the electrode surface. Therefore, despite excellent glucose-sensing properties, the electrodes are not usable for multiple tests. However, the economic aspect can be a good choice for single-use testing strips.

Since the real-time detection is the main objective, glucose detection is done using chronoamperometry. The glucose oxidation occurs at 0.55 V (reduction potential of Ni as seen in Figure 5 a)), so the chronoamperometric detection is done at 0.6 V. A stabilization period of 100 seconds was chosen to achieve a stable background response. Then with addition of 1 mM glucose, consequently at each addition, a 30 second analysis window is used for recording glucose sensing response. To optimize the device response, the first parameter adjusted was the loading of the nanomaterial. For this purpose, a known concentration of NiO NF dispersion is prepared, and 20 µL of it is drop cast onto the Ex-G electrodes. These concentrations vary from 0.5 - 3 mg/mL. The glucose sensitivity increases with an increase in loading concentration up to 3 mg/mL. The comparative plot of current response depending upon the different loading concentrations is shown in Figure 6 a). From this figure, it is evident that with increasing glucose concentrations the current response changes in step increments. These step increments have higher values for greater loading concentration. This is an important parameter for achieving the best glucose sensing response. Larger the change in current per unit glucose concentration, better the electrode sensitivity. We observed that the best response was recorded at 3 mg/mL loading concentration. Further increment in loading leads to a decrease in overall sensitivity. To show the comparative glucose sensitivity with respect to traditional glassy carbon electrodes, 3 mg/mL loaded GCE electrodes were also prepared, and their glucose responsivity was tested using the same electrochemical parameters. A comparative chronoamperometric response is presented in Figure 6 b). The stark difference in the glucose-sensing response per unit concentration is clear from this graph. However, this also brings forth one of the small hinderances, the large background current. The background current is one order higher than the GCE. This is most possibly due to the manual preparation of the electrode, which involves use of silver paint and thin copper wires for contact. However, the important parameter responsivity of Ex-G electrodes is also an order higher than that of GCE, negating the effects of higher background signals. The overall increase in surface area leads to more efficient electrooxidation and better charge transfer during the oxidation process. Thereby increasing the overall responsivity and sensitivity of the Ex-G electrodes compared to the standard GCE. The glucose sensing response of these electrodes is achieved even at concentrations >15 mM. However, the linear range is best within 1-10 mM glucose concentration. It is important to define the terms of merit of the glucose sensor, such as responsivity, sensitivity, and linear range, for a clear ideation of its performance. The responsivity of any sensor is defined as the change in current response per unit change in the stimuli. The stimuli in this case are the glucose concentration. The average responsivity of the 3 mg/mL coated Ex-G electrodes was found to be ~ 60 µA/mM. At the same time, the best responsivity recorded was 66.5 µA/mM. In contrast, the GCE showed a responsivity of 17.4 µA/mM.

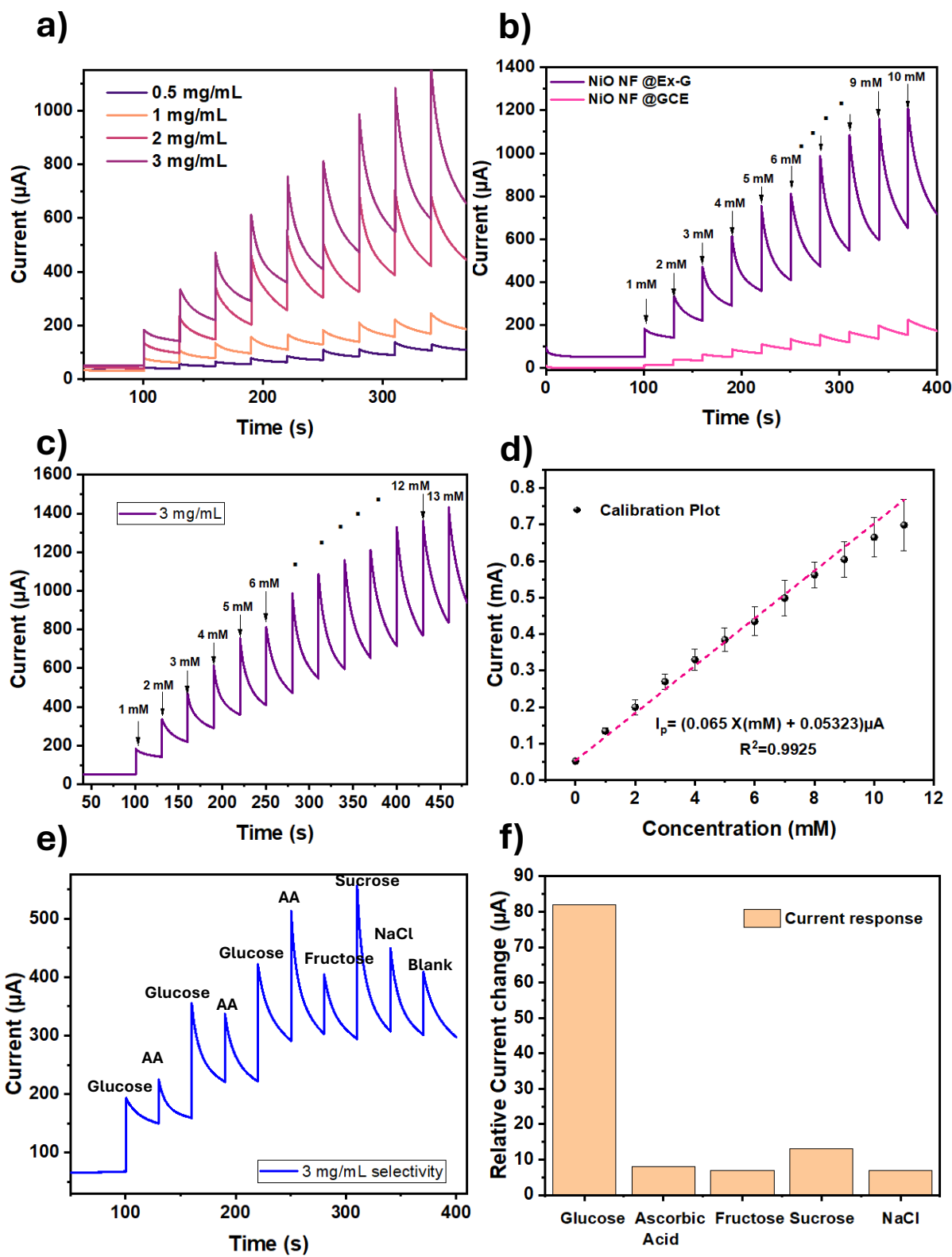

**Figure 6: a)** The chronoamperometric responses of NiO NF@Ex-G electrodes at various loading concentrations. **b)** The comparison of responses of 3 mg/mL NiO NF loaded GCE and Ex-G to increasing glucose concentration (1-8 mM). **c)** The chronoamperometric response of 3 mg/mL loaded Ex-G electrode with increasing glucose concentration from 1-13 mM. **d)** Calibration curve for the glucose sensing response corresponding to electrodes with 3 mg/mL NiO NF loading. **e)** The selectivity response during chronoamperometric detection of glucose against interfering agents such as Ascorbic Acid, fructose, sucrose and NaCl. **f)** The bar graph showing comparative current responses of interfering agents against Glucose as shown in chronoamperometric response in Figure 6 e).

This stark difference in responsivity is evident from the graph shown in Figure 6 b). The higher responsivity of the NiO NF decorated Ex-G electrodes can be attributed to two important factors. First, the greater geometrical surface area of the electrodes. The GCE used are of ~ 3 mm diameter, whereas the ExG electrodes are 5 mm in diameter, increasing the geometrical surface area by 1.8 times that of GCE. However, the responsivity of Ex-G electrodes is 3.5 times greater than that of the similarly prepared GCE. This additional difference is due to the second parameter, the higher electrochemically active surface area of the Ex-G electrodes. The more appropriate parameter for sensor performance that removes the effect of the geometric surface area is sensitivity. It is simply the responsivity per unit electrode area. The average sensitivity for the 3 mg/mL Ex-G electrodes is 304.12 µA/mM cm$^2$. The best sensitivity achieved is 339.28 µA/mM cm$^2$. In comparison, the same NiO NF loaded onto the GCE showed a 245.07 µA/mM cm$^2$ sensitivity. This corresponds to an increment of 24% in the sensitivity of the glucose sensing performance of the identical NiO NF particles. The final important parameter for judging sensor performance is the Limit of Detection (LOD). It is the lowest possible concentration of analyte that can be easily identified using a particular sensor. According to IUPAC, the LOD[33] is given by the formula :

$$LOD = (t \times SD\ of\ Blank)/m$$

Where, t = Students' T function

SD of Blank = Standard Deviation of the Blank Responses

m = Slope of the calibration curve.

Thus, to find out the value of these variables, five samples of 3 mg/mL loaded ExG electrodes were prepared, and their glucose response was tested by increment of 1 mM glucose. The calibration for current responses with respect to glucose concentrations was done by linear fitting of the response data achieved. The calibration plot is shown in Figure 6 d). The slope of the calibration plot was found to be 0.065 mA/mM. The fitting parameter R$^2$ for this linear fitting was 0.9925, showing a good linear response. Thus, the limit of detection of the sensor was found to be ~0.1 mM.

Finally, an important feature of any biosensor is its selectivity against other interfering agents present in the serum sample. This test was conducted by testing against common interfering agents such as Ascorbic Acid, Sucrose, Fructose, and NaCl. The chronoamperometric response for selectivity test is shown in Figure 6 e). Here, 1 mM glucose concentrations show substantial change in sensor current. However, with 0.1 mM addition of other analytes such as ascorbic acid, sucrose and fructose; there is negligible change in the chronoamperometric response. The bar graph shown in Figure 6 f) presents a more numerically visual picture of the excellent selectivity of the sensor. Here, despite the presence of other analytes, the sensor performance for glucose has shown less than 10 µA change in the current responses.

**Conclusion:**

Modern glucose-sensing disposable chips utilize age-old electrodes to determine the non-enzymatic glucose-sensing properties of metal oxide materials. The proposed Ex-G electrode offers a low-cost yet better-performing alternative to the standard GCE and SPCEs available in the literature. It shows an excellent glucose responsivity of ~ 60 µA/mM and a sensitivity of 304.12 µA/mM cm$^2$. This sensitivity indicates that the NiO NF decorated Ex-G electrodes are 24% better performing than the NiO NF decorated GCEs. The incremented sensitivity is complemented by high selectivity against common interference agents, most importantly against ascorbic acid, which is considered one of the most concerning analytes, present in serum samples. With suitable changes, the proposed sensor can be an alternative to the GCEs and the SPCEs in the near future.

**Acknowledgment:**

Author Ambesh Dixit acknowledges the Department of Science & Technology, Govt. of India through project # DST/TTI/BDTD/08/2024 and Ministry of Micro, Small & Medium Enterprises (MSME), Govt. of India through project # IDEARJ016782 for financial support to carry out this work. Vijay Kumar Singh acknowledges the Government of India for financial assistance. Authors also acknowledge the Center for Research and Development of Scientific Innovation (CRDS), IIT Jodhpur, for characterization facilities used in this work. Piyush Choudhary acknowledges the MoE, MSME, and the Government of India for funding and scholarship, respectively. All authors would like to acknowledge A-MAD research group members, IIT Jodhpur, for their support and help on this topic.